\documentstyle[osa,manuscript]{revtex}
\newcommand{\MF}{{\large{\manual META}\-{\manual FONT}}}
\newcommand{\manual}{rm}
\newcommand\bs{\char '134 }
\begin{document}

\title{Stress--softening and recovery of elastomers}
\author{Aleksey D. Drozdov$^{1}$ and Al Dorfmann$^{2}$}
\address{$^{1}$ Institute for Industrial Mathematics, 4 Hanachtom Street\\
84311 Beersheba, Israel\\
$^{2}$ Institute of Structural Engineering, 82 Peter Jordan Street\\
1190 Vienna, Austria}
\maketitle

\begin{abstract}
A constitutive model is developed for the mechanical response
of elastomers at finite strains.
A polymer is treated as a network of linear chains
linked by permanent (chemical crosslinks)
and temporary (entanglements and van der Waals forces) junctions.
Temporary junctions are assumed to be in two states:
loose (passive) when they impose only topological constrains
on available configurations of chains,
and tight (active) when their effect is tantamount to that for crosslinks.
Stretching of a specimen implies that some loose junctions become active,
which decreases the average length of a chain.

A long chain is treated as an ensemble of inextensible strands
connected in sequel.
Two neighboring strands are bridged by a bond which may be in two
conformations: flexed (trans) and extended (cis).
A bond in the flexed conformation is modeled as a linear elastic
solid, whereas the mechanical energy of a bond in the extended
conformation (two rigid rods directed along a straight line) is
disregarded.
For a virgin specimen, all bonds are in the flexed conformation.
Under loading some bonds are transformed from flexed to
extended conformation.

Stress--strain relations for a rubbery polymer and kinetic equations
for the trans--cis transition are derived using the laws of
thermodynamics.
Governing equations are determined by 5 adjustable parameters
which are found by fitting experimental data in uniaxial tensile
tests on natural rubber vulcanizates with various amounts of
crosslinks.
Fair agreement is demonstrated between results of numerical simulation
and observations with the elongation ratio up to $k=8$.
We analyze the effects of cyclic loading, thermal annealing and
recovery by swelling on the material constants.
\end{abstract}
\vspace*{3 mm}

{\bf Key--words:} Elastomers, Stress--softening, Recovery,
Rigid--rod chains, Temporary networks

\section{Introduction}

This study is concerned with modeling stress--softening
and subsequent recovery (by annealing at an elevated temperature
and by swelling) of elastomers.
Stress--strain relations for unfilled and particle reinforced
rubbery polymers have been the focus of attention
in the past four decades.
This may be explained by numerous applications of rubbery-like
materials in industry (vehicle tires, seals, shock absorbers,
flexible joints, etc.), as well as by some peculiarities
in their mechanical behavior whose physical mechanisms are not
quite clear.
A renewal of the interest to the mechanical response of elastomers
observed in the past decade, see Refs.
\cite{GS92,AB93,JB93,WG93,BB94,JB95,HS96,Lio96,BB98,SE98,OR99,BA00,BK00,MM00,MK00},
is associated with the development of constitutive equations
that (i) account for the molecular structure of polymers
at the micro-level
and (ii) describe some features of their response induced
by time-dependent changes in the internal structure
(viscoplasticity, damage, micro-fracture, etc.).

Among these peculiarities, one of the most important for applications
is the Mullins effect \cite{Mul47}.
This phenomenon is evidenced in tensile \cite{MT65,HMP65,HP66,HP67},
compressive \cite{BB98} and shear \cite{ES99}
tests as a noticeable difference between the stress--strain
curve for a virgin specimen and that for the material
reloaded after retraction.
A conventional standpoint is that cyclic pre-stretching with 3 to 6 cycles
and the maximum elongation ratio of about $k=2$ makes the response
of an elastomer repeatable \cite{BB98}.
However, the number of cycles, as well as their amplitude and frequency
are chosen by the trial-and-error method and substantially depend
on what the repeatability of a stress--strain diagram means for the
experimentalist.

Although the physical mechanism of the Mullins effect in unfilled
elastomers remains hitherto obscure, it is traditionally associated
with material damage \cite{GS92,OR99,MK00},
non-affinity of deformation for a polymeric network \cite{SE98,ES99},
viscous drag excerted on the network chains
by their environment \cite{Rol89}
and mechanically-induced crystallization of rubbery polymers \cite{HP67}.
These reasons may, however, be questioned, because observations
reveal that with an increase in strains
above the maximal level reached at pre-stretching,
the stress--strain curve returns to that for a virgin sample.

The present study aims to explain stress-softening of elastomers
under cyclic loading by an increase in the average size of
globules formed by polymeric chains (uncoiling of macromolecules).
For this purpose, we develop a phenomenological model for
the stress--strain response of rubbery polymers and determine
its parameters by fitting experimental data in uniaxial tensile
tests for virgin, pre-stretched and recovered specimens.

An elastomer is taken as a network of linear macromolecules
bridged by permanent (chemical crosslinks) and temporary
(physical crosslinks, entanglements and van der Waals forces)
junctions.
For the sake of simplicity, it is presumed that junctions
move affinely with the bulk material.
The novelty of our approach is that the number of junctions
is not fixed, but assumed to alter under loading.
To explain this phenomenon, we ascribe two possible states
to an entanglement: loose (passive) and tight (active).
An entanglement in the loose state does not create a junction,
but imposes only topological constrains on the number of available
configurations for a chain.
On the contrary, an entanglement in the tight state is thought of
as the same junction as that formed by a chemical crosslink.
The transition of an entanglement from its passive state to the
active state (tightening of entanglements) is supposed to be driven
by mechanical factors.
Introducing a simple kinetic equation for the rate of this transition,
we study the response of a temporary network with a time-varying
number of junctions.

A chain that connects two neighboring junctions is modeled
as a sequence of strands (statistically independent segments)
bridged by bonds \cite{BEV88}.
With reference to the Kratky--Porod concept \cite{BDM98},
strands are taken as inextensible (rigid) rods,
whereas the mechanical energy of a chain
is determined as the sum of strain energies for bonds.
It is postulated that a bond may be in two stable conformations:
flexed (trans) and extended (cis).
A bond in the flexed conformation is thought of as a linear elastic
solid, whereas the mechanical energy of a bond in the extended
conformation vanishes.
Constitutive equations for an elastomer and the kinetic equations
for the transition of bonds from their flexed to extended conformations
are derived by using the laws of thermodynamics.
The development of stress--strain relations is based on the
hypothesis that the characteristic time for changes in the concentration
of active entanglements substantially exceeds
that for mechanical deformation.

The paper is organized as follows.
Deformation of a chain is discussed in Section 2.
The mechanical energy of a temporary network is determined
in Section 3.
Stress--strain relations are developed in Section 4 using the
laws of thermodynamics.
Kinetic equations for transition of entanglements from their
loose to tight state are introduced in Section 5.
Uniaxial extension of a specimen is studied in Section 6.
Results of numerical simulation are compared with experimental
data for natural rubber vulcanizates in Section 7.
Section 8 deals with the effects of cyclic stretching and
recovery on experimental constants.
Some concluding remarks are formulated in Section 9.

\section{Deformation of a long chain}

An elastomer is modeled as an ensemble of linear macromolecules
bridged by crosslinks, entanglements and van der Waals forces.
Active links between macromolecules are thought of as junctions
(permanent in the case of chemical crosslinks
and temporary for entanglements and van der Waals forces).
A sequence of mers (belonging to a polymeric molecule)
between two subsequent junctions is associated with a chain.
Chains are divided into strands (statistically
independent segments) bridged by bonds.
In accord with the Kratky--Porod model,
a chain is treated as an aggregate consisting
of $N+1$ identical inextensible strands linked in sequel.
The average number of bonds in a chain change under stretching
of a specimen because of slippage of entanglements with respect
to chains and transition of entanglements from their loose
to the tight state.
This implies that $N$ is taken as a function of time, $N=N(t)$.

We adopt a zigzag model \cite{Rob66} and assume that
a bond bridging two neighboring strands is characterized
by one of the two stable conformations: flexed and extended.
In the stress-free state the angle between two strands
linked by a bond in the flexed conformation equals
$\theta\in (0,\pi)$ (the same for all bonds)
and that for a bond in the extended conformation equals
$\pi$ (see a sketch depicted in Figure~1 of Ref. \cite{JM90}).
In thermal equilibrium before deformation all bonds
are in the flexed conformation.
Mechanical loading activates chains, which implies that
conformations of some bonds alter.
The numbers of bonds in flexed and extended conformations,
$N_{\rm f}(t)$ and $N_{\rm e}(t)$, obey the balance law
\[ N_{\rm f}(t)+N_{\rm e}(t)=N(t). \]
Introducing the average concentration of bonds in the extended
conformation, $n=n(t)$, we obtain
\begin{equation}
N_{\rm f}(t)=N(t) [1-n(t)],
\qquad
N_{\rm e}(t)=N(t) n(t).
\end{equation}
In a deformed state the angle between strands linked by a bond in
the flexed conformation changes, whereas for a bond in
the extended conformation
(modeled as two rigid rods directed along a straight line)
this angle remains unaltered.
The strain, $e$, from the stress-free state of a chain
to its deformed state equals the sum of strains
for bonds in the flexed conformation,
\[ e=N_{\rm f}e_{\rm f}. \]
This equality together with Eq. (1) implies that
\begin{equation}
e_{\rm f}=\frac{e}{N(1-n)}.
\end{equation}
Bonds in the flexed conformation are modeled as linear elastic solids
with the mechanical energy
\[ \frac{1}{2} \mu e_{\rm f}^{2}, \]
whereas the strain energy of bonds in the extended conformation vanishes.
The mechanical energy of a chain, $w$, equals the sum
of the mechanical energies for individual bonds.
It follows from Eqs. (1) and (2) that
\begin{equation}
w(e) =\frac{\mu e^{2}}{2 N(1-n)}.
\end{equation}
To express the strain in a chain, $e$, in terms of
the macro-strain tensor for a network, $\hat{\epsilon}$,
we consider a chain connecting neighboring junctions
$A_{1}(t)$ and $A_{2}(t)$.
Let $\bar{r}_{01}(t)$ and $\bar{r}_{02}(t)$ be radius vectors
of these points in the stress--free state and
$\bar{r}_{1}(t)$ and $\bar{r}_{2}(t)$ their radius vectors
in the deformed state at time $t\geq 0$.
The end-to-end length of the chain reads $\delta_{0}(t)$
in the stress-free state and $\delta(t)$ in the deformed state.
Introducing the guiding vector in the stress-free state, $\bar{l}$,
(the unit vector directed along the end-to-end vector for the chain),
we obtain
\[ \bar{r}_{02}(t)-\bar{r}_{01}(t)=\delta_{0}(t) \bar{l}. \]
In the deformed state, the junctions occupy points with
the radius vectors
\[ \bar{r}_{1}(t)=\bar{r}_{01}(t)+\bar{u}(t,\bar{r}_{01}(t)),
\qquad
\bar{r}_{2}(t)=\bar{r}_{02}(t)+\bar{u}(t,\bar{r}_{02}(t)), \]
where $\bar{u}(t,\bar{r})$ is the displacement vector
at point $\bar{r}$ for transition from the stress-free state
to the deformed state at time $t$.
The end-to-end vector for the chain in the deformed state is given by
\[ \bar{R}(t) = \bar{r}_{2}(t)-\bar{r}_{1}(t)
= \delta_{0}(t)\bar{l}+[\bar{u}(t,\bar{r}_{01}(t)+\delta_{0}(t)\bar{l})
-\bar{u}(t,\bar{r}_{01}(t))]. \]
Neglecting terms beyond the first order of smallness compared
to $\delta_{0}(t)$, we find that
\[ \bar{R}(t)=\delta_{0}(t) \bar{l} \cdot [\hat{I}
+\bar{\nabla}_{0}\bar{u}(t)], \]
where $\bar{\nabla}_{0}$ is the gradient operator
in the stress-free state,
$\hat{I}$ is the unit tensor,
the dot stands for inner product
and the argument $\bar{r}_{01}$ is omitted.
In terms of the radius vector in the deformed state, $\bar{r}$,
this equality reads
\begin{equation}
\bar{R}(t)=\delta_{0}(t) \bar{l}\cdot \bar{\nabla}_{0}\bar{r}(t)
=\delta_{0}(t) [ \bar{\nabla}_{0}\bar{r}(t) ]^{\top} \cdot \bar{l},
\end{equation}
where $\top$ stands for transpose.
The end-to-end length of the chain in the deformed state, $\delta(t)$,
is given by
\[ \delta^{2}=\bar{R}\cdot \bar{R}. \]
This equality together with Eq. (4) yields
\begin{equation}
\delta^{2}(t,\bar{l})
= \delta_{0}^{2}(t) \bar{l}\cdot \hat{C}(t)\cdot \bar{l},
\end{equation}
where
\begin{equation}
\hat{C}(t)=\bar{\nabla}_{0}\bar{r}(t)
\cdot [ \bar{\nabla}_{0}\bar{r}(t) ]^{\top}
\end{equation}
is the Cauchy deformation tensor for transition from
the stress-free state of the network to its deformed
state at time $t$.

The extension ratio for the chain, $\lambda(t,\bar{l})$,
is defined as the ratio
of the current end-to-end length, $\delta(t,\bar{l})$, to that of the chain
in its activated stress-free state, $\delta^{\circ}(t,\bar{l})$.
The latter state is defined as a state in which the chain is
unloaded, but the numbers of bonds in various conformations coincide
with their current values, $N_{\rm f}(t,\bar{l})$ and $N_{\rm e}(t,\bar{l})$.
It differs from the equilibrium stress-free state of the chain,
where the number of bonds in the extended conformation vanishes.
The difference between the end-to-end lengths of a chain
in the activated state, $\delta^{\circ}(t,\bar{l})$,
and in the equilibrium state, $\delta_{0}(t)$, determines
the end-to-end elongation
driven by transformation of bonds from their flexed conformation
to the extended conformation.
We assume that the transformation-induced end-to-end elongation
is proportional to the number of bonds acquiring the extended
conformation,
\begin{equation}
\delta^{\circ}(t,\bar{l})=\delta_{0}(t)[1+\eta n(t,\bar{l})],
\end{equation}
where $\eta>0$ characterizes an increment
of the end-to-end length driven by an individual transition.
The quantity $\eta$ reflects the average size of a globule
created by a polymeric chain:
when $\eta$ is small, the chain is rolled into a tight coil,
whereas an increase in $\eta$ is tantamount to unfolding the coil.
Because the parameter $\eta$ describes the current state of
a chain, it should be thought of as a function of time (strain).
To simplify calculations, we treat it as a constant, but suppose
that $\eta$ can change from one stress--strain curve to another
in cyclic tests.

It follows from Eqs. (5) and (7) that
the Hencky strain for a chain, $e=\ln \lambda$, reads
\begin{equation}
e(t,\bar{l}) = \ln \frac{\delta(t,\bar{l})}{\delta^{\circ}(t,\bar{l})}
=\frac{1}{2} \ln \Bigl [ \bar{l}\cdot\hat{C}(t)\cdot\bar{l}\Bigr ]
-\ln \Bigl [ 1+\eta n(t,\bar{l})\Bigr ].
\end{equation}
Equations (3) and (8) determine the mechanical energy of a chain, $w$,
in terms of the Cauchy deformation tensor for the network, $\hat{C}$.

\section{Strain energy density of a network}

We adopt the conventional hypothesis that the excluded-volume effect
and other multi-chain effects are screened for an individual chain
by surrounding macromolecules \cite{Eve98}.
This implies that the energy of interaction between chains
is neglected (under the incompressibility condition for the network)
and the mechanical energy of the network equals the sum
of the strain energies for individual chains.
Assuming the distribution of chains to be isotropic,
we find the concentration of chains in
a network (per unit mass) with guiding vector $\bar{l}$,
\[ X(t,\bar{l})=\frac{\Xi(t)}{4\pi} \sin\vartheta d\vartheta d\varphi , \]
where $\vartheta$ and $\varphi$ are Euler's angles
which determine the position of the unit vector $\bar{l}$
with respect to some Cartesian coordinate frame
and $\Xi(t)$ is the number of chains (per unit mass)
at time $t\geq 0$.
To determine the strain energy of a network,
we multiply the number of chains, $X(t,\bar{l})$,
by their mechanical energy, $w$,
and sum the results for various guiding vectors, $\bar{l}$,
\begin{equation}
W(t) = \frac{\mu \Xi(t)}{8\pi N(t)} \int_{0}^{2\pi} d\varphi
\int_{0}^{\pi} \frac{e^{2}(t,\vartheta,\varphi)}{1
-n(t,\vartheta,\varphi)} \sin\vartheta d\vartheta.
\end{equation}
The quantities $N(t)$ and $\Xi(t)$ are connected by the mass
conservation law
\begin{equation}
m N(t)\Xi(t)=1,
\end{equation}
where $m$ stands for the average mass of a strand.
It follows from Eqs. (9) and (10) that
\begin{equation}
W(t) = \frac{\mu m \Xi^{2}(t)}{8\pi} \int_{0}^{2\pi} d\varphi
\int_{0}^{\pi} \frac{e^{2}(t,\vartheta,\varphi)}{1
-n(t,\vartheta,\varphi)} \sin\vartheta d\vartheta.
\end{equation}
Let $\Xi_{\ast}$ be the average concentration of chains (per unit mass)
in a (hypothetical) totally disentangled network and $M(t)$ the
current number of active entanglements per unit mass.
Because any tight entanglement doubles the number of chains
involved in the knot,
the concentration of chains in the entangled network reads
\begin{equation}
\Xi(t)=\Xi_{\ast}+2M(t)=\Xi_{0}[1+\nu(t)],
\end{equation}
where $\Xi_{0}=\Xi_{\ast}+2M(0)$ is the average number of chains
(per unit mass) in an equilibrium network (before a test)
and the dimensionless function
\begin{equation}
\nu(t)=2\frac{M(t)-M(0)}{\Xi_{0}}
\end{equation}
characterizes the ratio of the number of loose
entanglements that have transformed into the active state
within the interval $[0,t]$ to the initial number of chains.
Combining Eqs. (11) and (12), we obtain
\begin{equation}
W(t) = \frac{\mu m \Xi_{0}^{2}}{8\pi} [1+\nu(t)]^{2}
\int_{0}^{2\pi} d\varphi
\int_{0}^{\pi} \frac{e^{2}(t,\vartheta,\varphi)}{1
-n(t,\vartheta,\varphi)} \sin\vartheta d\vartheta.
\end{equation}
The average mechanical energy per chain is given by
\[ w_{0}(t)=\frac{W(t)}{\Xi(t)}. \]
Substitution of Eqs. (12) and (14) into this equality yields
\begin{equation}
w_{0}(t) = \frac{\mu m \Xi_{0}}{8\pi} [1+\nu(t)]
\int_{0}^{2\pi} d\varphi
\int_{0}^{\pi} \frac{e^{2}(t,\vartheta,\varphi)}{1
-n(t,\vartheta,\varphi)} \sin\vartheta d\vartheta.
\end{equation}

Our objective now is to calculate the derivative of the function $W$
with respect to time.
For this purpose, we suppose that the characteristic rate for
transition of bonds from their flexed to extended conformation
substantially exceeds that for changes in the concentration of
active (tight) entanglements.
This assumption may be explained by the fact that the characteristic
length of a chain (the length between neighboring entanglements)
dramatically exceeds that for a strand (the length-scale
associated with a bond).
This implies that the function $\nu(t)$ may be treated as a
constant when expression (14) is differentiated with respect to $t$.
Bearing this hypothesis in mind, we find from Eq. (14) that
\begin{eqnarray}
\frac{dW}{dt}(t) &=& \frac{\mu m\Xi_{0}^{2}}{4\pi}[1+\nu(t)]^{2}
\int_{0}^{2\pi} d\varphi \int_{0}^{\pi}
\frac{e(t,\vartheta,\varphi)}{1-n(t,\vartheta,\varphi)}
\frac{\partial e}{\partial t}(t,\vartheta,\varphi)
\sin\vartheta d\vartheta +J_{1}(t),
\nonumber\\
J_{1}(t) &=& \frac{\mu m\Xi_{0}^{2}}{8\pi} [1+\nu(t)]^{2}
\int_{0}^{2\pi} d\varphi \int_{0}^{\pi}
\frac{e^{2}(t,\vartheta,\varphi)}{[1-n(t,\vartheta,\varphi)]^{2}}
\frac{\partial n}{\partial t}(t,\vartheta,\varphi)
\sin\vartheta d\vartheta.
\end{eqnarray}
It follows from Eq. (8) that
\begin{equation}
\frac{\partial e}{\partial t}(t,\bar{l}) =
\frac{1}{2[\bar{l}\cdot\hat{C}(t)\cdot\bar{l}]}
\Bigl [\bar{l}\cdot\frac{d\hat{C}}{dt}(t)\cdot\bar{l}\Bigr ]
-\frac{\eta}{1+\eta n(t,\bar{l})}
\frac{\partial n}{\partial t}(t,\bar{l}).
\end{equation}
Differentiation of Eq. (6) with respect to time yields
\[
\frac{d\hat{C}}{dt}(t)=\bar{\nabla}_{0}\bar{v}(t)\cdot
[\bar{\nabla}_{0}\bar{r}(t)]^{\top}
+\bar{\nabla}_{0}\bar{r}(t)\cdot [\bar{\nabla}_{0}\bar{v}(t)]^{\top},
\]
where
$\bar{v}(t)=d\bar{r}(t)/dt$ is the velocity vector for the network.
Taking into account that
\[ \bar{\nabla}_{0}\bar{v}(t)=\bar{\nabla}_{0}\bar{r}(t)
\cdot \bar{\nabla}(t)\bar{v}(t), \]
where $\bar{\nabla}(t)$ is the gradient operator in the deformed state
at time $t$, we obtain
\begin{equation}
\frac{d\hat{C}}{dt}(t)=2\bar{\nabla}_{0}\bar{r}(t)\cdot\hat{D}(t)
\cdot [\bar{\nabla}_{0}\bar{r}(t)]^{\top},
\end{equation}
where
\[ \hat{D}(t)=\frac{1}{2}\Bigl [\bar{\nabla}(t)\bar{v}(t)
+(\bar{\nabla}(t)\bar{v}(t))^{\top}\Bigr ] \]
is the rate-of-strain tensor for the network.
It follows from Eq. (18) that
\begin{equation}
\bar{l}\cdot \frac{d\hat{C}}{dt}(t)\cdot\bar{l}
= 2\bar{l}\cdot \bar{\nabla}_{0}\bar{r}(t)\cdot\hat{D}(t)
\cdot [\bar{\nabla}_{0}\bar{r}(t)]^{\top}\cdot \bar{l}
=2\hat{F}(t,\bar{l}):\hat{D}(t),
\end{equation}
where 
\begin{equation}
\hat{F}(t,\bar{l})=[\bar{\nabla}_{0}\bar{r}(t)]^{\top}\cdot
(\bar{l}\otimes \bar{l}) \cdot \bar{\nabla}_{0}\bar{r}(t)
\end{equation}
is the generalized Finger tensor \cite{Dro00},
the colon stands for convolution
and $\otimes$ denotes tensor product.
Substitution of Eqs. (17) and (19) into Eq. (16) implies that
\begin{eqnarray}
&& \frac{dW}{dt}(t) = \hat{\Upsilon}(t):\hat{D}(t)-J(t),
\nonumber\\
&& \hat{\Upsilon}(t) = \frac{\mu m\Xi_{0}^{2}}{4\pi}[1+\nu(t)]^{2}
\int_{0}^{2\pi} d\varphi \int_{0}^{\pi}
\frac{e(t,\vartheta,\varphi)}{1-n(t,\vartheta,\varphi)}
\frac{\hat{F}(t,\vartheta,\varphi)}{\bar{l}\cdot\hat{C}(t)\cdot\bar{l}}
\sin\vartheta d\vartheta,
\nonumber\\
&& J(t) = \frac{\mu m\Xi_{0}^{2}}{8\pi}[1+\nu(t)]^{2}
\int_{0}^{2\pi} d\varphi \int_{0}^{\pi}
\frac{H(t,\vartheta,\varphi)}{[1-n(t,\vartheta,\varphi)]^{2}}
\frac{\partial n}{\partial t}(t,\vartheta,\varphi)
\sin\vartheta d\vartheta,
\nonumber\\
&& H(t,\vartheta,\varphi) = e(t,\vartheta,\varphi)
\biggl \{ \frac{2\eta [1-n(t,\vartheta,\varphi)]}
{1+\eta n(t,\vartheta,\varphi)}
-e(t,\vartheta,\varphi) \biggr \}.
\end{eqnarray}

\section{Constitutive equations}

Observations evidence that under cyclic loading with small frequency
(less than 10 Hz), the temperature increment is negligible
and temperature $T$ remains close to its reference value $T_{0}$
\cite{KB97}.
This means that the effect of temperature on material parameters,
as well as thermal expansion of the network may be disregarded.
It is assumed that the deformation process is rather slow,
which implies that at any instant $t\geq0$ thermodynamic potentials
are correctly defined.
For affine deformation of an incompressible network,
the Clausius--Duhem inequality reads \cite{CG67}
\begin{equation}
T \frac{dQ}{dt}=-S\frac{dT}{dt}-\frac{d\Psi}{dt}
+\frac{1}{\rho}\Bigl ( \hat{\sigma}_{\rm d}:\hat{D}
-\frac{1}{T}\bar{q}\cdot \bar{\nabla} T \Bigr) \geq 0.
\end{equation}
where $\rho$ is mass density,
$\bar{q}$ is the heat flux vector,
$\hat{\sigma}_{\rm d}$ is the deviatoric component
of the Cauchy stress tensor $\hat{\sigma}$,
$\Psi$ is the free (Helmholtz) energy,
$S$ is the entropy
and $Q$ is the entropy production per unit mass.
We accept the following expression for the free energy:
\begin{equation}
\Psi = \Psi_{0}+ (c-S_{0})(T-T_{0}) -cT\ln\frac{T}{T_{0}}+W,
\end{equation}
where $S_{0}$ and $\Psi_{0}$ are the entropy and the free energy
in the equilibrium stress-free state at the reference temperature $T_{0}$
and $c$ is the specific heat.
The second and third terms on the right-hand side of Eq. (23)
characterize the energy of thermal motion.
Unlike conventional theories of rubber elasticity \cite{Tre75},
the terms associated with configurational entropy of chains
are disregarded in Eq. (23).
According to Ref. \cite{JM90}, this approximation is acceptable,
provided that
\[ \mu\gg k_{B}TN, \]
where $k_{B}$ is Boltzmann's constant.
In the sequel, we suppose that this inequality is satisfied.

Substituting Eqs. (14) and (23) into Eq. (22) and assuming changes
in the concentration of junctions to be rather slow
(which means that Eq. (21) for the derivative of
the mechanical energy $W$ with respect to time may be used),
we arrive at the formula
\begin{equation}
T\frac{dQ}{dt} = \biggl (\frac{\hat{\sigma}_{\rm d}}{\rho}
-\hat{\Upsilon}\biggr ):\hat{D}
-\biggl ( S-S_{0}-c\ln\frac{T}{T_{0}}\biggr )\frac{dT}{dt}
+J-\frac{1}{\rho T}\bar{q}\cdot\bar{\nabla}T \geq 0.
\end{equation}
Applying the conventional reasoning \cite{CG67} to Eq. (24),
we find that the expressions in braces vanish.
This assertion results in the standard formula for the entropy
\begin{equation}
S=S_{0}+c\ln\frac{T}{T_{0}},
\end{equation}
where the configurational entropy of the network is neglected,
and, together with Eqs. (20) and (21), the constitutive equation
\begin{equation}
\hat{\sigma}(t) = -P(t)\hat{I}
+G [1+\nu(t)]^{2}[\bar{\nabla}_{0}\bar{r}(t)]^{\top}
\cdot \int_{0}^{2\pi} d\varphi \int_{0}^{\pi}
\frac{e(t,\vartheta,\varphi)}{1-n(t,\vartheta,\varphi)}
\frac{\bar{l}\otimes \bar{l}}{\bar{l} \cdot\hat{C}(t)\cdot \bar{l}}
\sin\vartheta d\vartheta\;\cdot \bar{\nabla}_{0}\bar{r}(t),
\end{equation}
where $P(t)$ is pressure and $G=\rho \mu m\Xi_{0}^{2}/(4\pi)$.
We substitute Eqs. (25) and (26) into Eq. (24) and find that the rate
of entropy production is nonnegative for an arbitrary loading program,
provided that
\begin{enumerate}
\item
the heat flux vector $\bar{q}$ obeys the Fourier law
$\bar{q}=-\kappa \bar{\nabla}T$ with a positive thermal
diffusivity $\kappa$,

\item
the function $n(t,\vartheta,\varphi)$ satisfies the kinetic equation
\begin{equation}
\frac{\partial n}{\partial t}(t,\vartheta,\varphi)
= \alpha(t) e(t,\vartheta,\varphi)
\biggl \{ \frac{2\eta [1-n(t,\vartheta,\varphi)]}
{1+\eta n(t,\vartheta,\varphi)} -e(t,\vartheta,\varphi)\biggr \},
\qquad
n(0,\vartheta,\varphi)=0,
\end{equation}
where $\alpha$ is a nonnegative function of time.
\end{enumerate}
We postulate that the rate of trans--cis transformation is
proportional to the average number of strands in a chain,
\begin{equation}
\alpha(t)=\alpha_{1} N(t),
\end{equation}
where $\alpha_{1}>0$ is a material constant.
Equation (28) is explained by the fact that the force opposing
transition of a bond from its flexed to extended conformation is
driven by the action of neighboring bonds,
whereas stresses in these bonds are inversely proportional
to the number of strands in a chain, see Eq. (3).
Combining Eqs. (27) and (28) and using Eqs. (10) and (12), we arrive at
the nonlinear differential equation
\begin{equation}
\frac{\partial n}{\partial t}(t,\vartheta,\varphi)
= \alpha_{0}\frac{e(t,\vartheta,\varphi)}{1+\nu(t)}
\biggl \{ \frac{2\eta [1-n(t,\vartheta,\varphi)]}
{1+\eta n(t,\vartheta,\varphi)} -e(t,\vartheta,\varphi)\biggr \},
\qquad
n(0,\vartheta,\varphi)=0
\end{equation}
with $\alpha_{0}=\alpha_{1}/(m\Xi_{0})$.

\section{Evolution of the concentration of junctions}

To close the model, the kinetics of stress--induced changes
in the concentration of temporary junctions should be described.
Denote by $M_{0}$ the total concentration (per unit mass)
of entanglements (both loose and tight).
Adopting the first-order kinetics for the mechanically-induced
evolution of the number of tight entanglements, we assume that the
rate of increase in the concentration of active entanglements
is proportional to the current number of loose ones,
\begin{equation}
\frac{dM}{dt}(t)=\beta_{\ast}(t)[M_{0}-M(t)],
\end{equation}
where $\beta_{\ast}(t)$ is a positive function.
It follows from Eqs. (12) and (30) that
\begin{equation}
\frac{d\nu}{dt}(t)=\beta_{\ast}(t)[\nu_{0}-\nu(t)],
\end{equation}
where the quantity
\[ \nu_{0}=2\frac{M_{0}-M(0)}{\Xi_{0}} \]
equals the ratio of the initial number of loose entanglements
to the initial number of chains.
We postulate that the rate of growth for the number of
active entanglements, $\beta_{\ast}$, is proportional
to the mechanical energy per chain and inversely proportional
to the number of strands in a chain,
\begin{equation}
\beta_{\ast}(t)=\beta_{0}\frac{w_{0}(t)}{N(t)},
\end{equation}
where $\beta_{0}>0$ is a material constant.
The proportionality of $\beta_{\ast}$ to the average mechanical
energy, $w_{0}$, reflects the fact that tightening
of loose entanglements is driven by mechanical factors.
In general, any measure of straining may be chosen instead of $w_{0}$
on the right-hand side of Eq. (31), see a discussion of this issue
in \cite{BK00,MK00}.
We introduce the strain energy density, $w_{0}$, by analogy
with conventional models for damage of elastomers \cite{Mie95}.
The inverse proportionality of $\beta_{\ast}$ to the average number
of strands in a chain, $N$, reflects slowing down of the process
of tightening entanglements with an increase in the
average length of chains.
The latter is driven by the growth of the chains' mobility
(estimated in terms of the number of available configurations).

It follows from Eqs. (10), (12), (31) and (32) that the function $\nu(t)$
obeys the kinetic equation
\begin{equation}
\frac{d\nu}{dt}(t)=\frac{\beta}{4\pi} [\nu_{0}-\nu(t)]
\int_{0}^{2\pi} d\varphi
\int_{0}^{\pi} \frac{e^{2}(t,\vartheta,\varphi)}{1
-n(t,\vartheta,\varphi)} \sin\vartheta d\vartheta,
\qquad
\nu(0)=0,
\end{equation}
where $\beta=\beta_{0}\mu/2$.
Governing equations (26), (29) and (33) are determined by 5 adjustable
parameters:
an analog of the shear modulus $G$,
the rate of trans--cis transition $\alpha_{0}$,
the rate of stress--induced increase in the concentration of active
entanglements $\beta$,
the constant $\eta$ which characterizes the end-to-end elongation
of a chain driven by transformation of a bond from its flexed to
extended conformation,
and the ratio, $\nu_{0}$, of the initial number
of loose entanglements to the initial number of chains.
To determine these quantities, we analyze uniaxial extension of a rod.

\section{Uniaxial tension of a specimen}

At uniaxial tension of an incompressible medium,
Cartesian coordinates in the deformed state, $x_{i}$,
are expressed in terms of the Cartesian coordinates
in the stress-free state, $X_{i}$, by the formulas
\[ x_{1}=k(t)X_{1},
\qquad
x_{2}=k^{-\frac{1}{2}}(t) X_{2},
\qquad
x_{3}=k^{-\frac{1}{2}}(t) X_{3}, \]
where $k=k(t)$ is the extension ratio.
It follows from these equalities and Eq. (6) that
\begin{eqnarray}
\bar{\nabla}_{0}\bar{r}(t) &=& k(t)\bar{e}_{1}\bar{e}_{1}
+k^{-\frac{1}{2}}(t) (\bar{e}_{2}\bar{e}_{2} +\bar{e}_{3}\bar{e}_{3}),
\nonumber\\
\hat{C}(t) &=& k^{2}(t)\bar{e}_{1}\bar{e}_{1}
+k^{-1}(t) (\bar{e}_{2}\bar{e}_{2} +\bar{e}_{3}\bar{e}_{3}).
\end{eqnarray}
The unit vector $\bar{l}$ is given by
\begin{equation}
\bar{l}=\cos\vartheta \bar{e}_{1}+
\sin\vartheta (\cos\varphi \bar{e}_{2}+\sin\varphi \bar{e}_{3} ),
\end{equation}
which implies that the tensor $\bar{l}\otimes \bar{l}$
is determined by the matrix
\begin{equation}
\bar{l}\otimes\bar{l}=\left [\begin{array}{ccc}
\cos^{2}\vartheta & \sin\vartheta\cos\vartheta\cos\varphi &
\sin\vartheta\cos\vartheta\sin\varphi\\
\sin\vartheta\cos\vartheta\cos\varphi & \sin^{2}\vartheta\cos^{2}\varphi &
\sin^{2}\vartheta\sin\varphi\cos\varphi\\
\sin\vartheta\cos\vartheta\sin\varphi & \sin^{2}\vartheta\sin\varphi\cos\varphi
& \sin^{2}\vartheta\sin^{2}\varphi
\end{array}\right ].
\end{equation}
Equations (34) and (35) yield
\begin{equation}
\bar{l}\cdot\hat{C}(t)\cdot\bar{l}=
k^{2}(t)\cos^{2}\vartheta+k^{-1}(t)\sin^{2}\vartheta.
\end{equation}
Because of the axial symmetry of deformation,
the functions $n=n(t,\vartheta)$ and $e=e(t,\vartheta)$
are independent of $\varphi$,
which implies that all terms but $\bar{l}\otimes \bar{l}$
on the right-hand side of Eq. (26) are independent of $\varphi$.
It follows from Eq. (36) that
\[ \frac{1}{2\pi} \int_{0}^{2\pi} \bar{l}\otimes \bar{l} d\varphi
=\cos^{2}\vartheta \bar{e}_{1}\bar{e}_{1}
+\frac{1}{2}\sin^{2}\vartheta (\bar{e}_{2}\bar{e}_{2}
+\bar{e}_{3}\bar{e}_{3} ). \]
Substitution of this expression and Eqs. (34) and (35)
into Eq. (26) implies that
\[ \hat{\sigma}(t)=\sigma(t)\bar{e}_{1}\bar{e}_{1}
+\sigma_{0}(t)(\bar{e}_{2}\bar{e}_{2}+\bar{e}_{3}\bar{e}_{3}), \]
where
\begin{eqnarray}
\sigma(t) &=& -P(t)+2\pi G [1+\nu(t)]^{2}
\int_{0}^{\pi} \frac{k^{2}(t)\cos^{2}\vartheta}{k^{2}(t)\cos^{2}\vartheta
+k^{-1}(t)\sin^{2}\vartheta}
\frac{e(t,\vartheta)}{1-n(t,\vartheta)}\sin\vartheta d\vartheta,
\nonumber\\
\sigma_{0}(t) &=& -P(t)+\pi G [1+\nu(t)]^{2}
\int_{0}^{\pi} \frac{k^{-1}(t)\sin^{2}\vartheta}{k^{2}(t)\cos^{2}\vartheta
+k^{-1}(t)\sin^{2}\vartheta}
\frac{e(t,\vartheta)}{1-n(t,\vartheta)}\sin\vartheta d\vartheta.
\end{eqnarray}
The boundary condition on the lateral surface of
the specimen implies that $\sigma_{0}(t)=0$.
Combining this equality with Eq. (38), we find the longitudinal stress
\[ \sigma(t) = \pi G [1+\nu(t)]^{2}
\int_{0}^{\pi} \frac{2k^{2}(t)\cos^{2}\vartheta
-k^{-1}(t)\sin^{2}\vartheta}{k^{2}(t)\cos^{2}\vartheta
+k^{-1}(t)\sin^{2}\vartheta}
\frac{e(t,\vartheta)}{1-n(t,\vartheta)}\sin\vartheta d\vartheta. \]
Introducing the notation
\[ z=\cos\vartheta,
\qquad
\tilde{e}(t,z)=e(t,\vartheta),
\qquad
\tilde{n}(t,z)=n(t,\vartheta) \]
and bearing in mind that $\tilde{e}$ and $\tilde{n}$
are even functions of $z$, we obtain
\begin{equation}
\sigma(t) = E [1+\nu(t)]^{2}
\int_{0}^{1}
\frac{2k^{2}(t)z^{2}-k^{-1}(t)(1-z^{2})}{k^{2}(t)z^{2}+k^{-1}(t)(1-z^{2})}
\frac{\tilde{e}(t,z)}{1-\tilde{n}(t,z)}dz,
\end{equation}
where
\begin{equation}
E=\frac{1}{2}\rho\mu m\Xi_{0}^{2}.
\end{equation}
It follows from Eqs. (8), (34) and (35) that
\begin{equation}
\tilde{e}(t,z) = \ln \frac{[k^{2}(t)z^{2}+k^{-1}(t)(1-z^{2})]^\frac{1}{2}}
{1+\eta \tilde{n}(t,z)}.
\end{equation}
In the sequel, we focus on stretching with a constant rate of
engineering strain, $\dot{\epsilon}_{0}>0$,
\begin{equation}
k(t)=1 + \dot{\epsilon}_{0}t.
\end{equation}
Combining Eqs. (29) and (42), we arrive at the kinetic equation
\begin{equation}
\frac{\partial \tilde{n}}{\partial k} = a\Bigl [
\frac{2\eta(1-\tilde{n})}{1+\eta\tilde{n}}-\tilde{e}\Bigr ]\tilde{e},
\qquad
\tilde{n}(1,z)=0
\end{equation}
where $a=\alpha/\dot{\epsilon}_{0}$.
Substituting expression (42) into Eq. (33) and calculating the
integral over $\varphi$, we find that
\begin{equation}
\frac{d\nu}{dk} = b (\nu_{0}-\nu) \int_{0}^{1}\frac{\tilde{e}^{2}dz}{1
-\tilde{n}},
\qquad
\nu(1)=0,
\end{equation}
where $b=\beta/\dot{\epsilon}_{0}$.
Equations (39), (41), (43) and (44) are determined by 5 adjustable
parameters: $E$, $a$, $b$, $\eta$ and $\nu_{0}$.
This number is quite comparable with the number of experimental constants
employed in conventional stress--strain relations in finite elasticity
of elastomers \cite{AB93,WG93,MM00,BDE81}.

\section{Comparison with experimental data}

To determine adjustable parameters, we fit observations for natural
rubber vulcanizates in uniaxial tensile tests with the strain rate
$\dot{\epsilon}_{0}=2.0$ min$^{-1}$ at room temperature.
For a description of specimens and the experimental procedure,
see Refs. \cite{HP66,HP67}.

We begin by matching experimental data in cyclic tests with
the maximal elongation $k_{\max}=6.0$ for a vulcanizate with an
unspecified composition.
First, we approximate the stress--strain curve for a virgin sample.
Given a Young's modulus, $E$, the constants $a$, $b$, $\eta$ and $\nu_{0}$
are found by the steepest-descent algorithm.
The parameter $E$ is determined by the least-squares technique.
Afterwards, we fix the values $b$ and $\nu_{0}$
(which are responsible for the evolution of the concentration
of junctions) and match stress--strain curves measured after retraction
by using only three constants: $E$, $a$ and $\eta$.
Experimental data in tests with various numbers of cycles, $i$,
are depicted in Figure~1 together with results of numerical simulation.
Material constants are listed in Table~1.
The parameters $E$, $a$ and $\eta$ are plotted versus the number of
cycles $i$ in Figures~2 and 3, which show that observations
are fairly well approximated by the stretched exponential functions
\begin{equation}
E=E_{0}\cdot 10^{i^{\kappa_{E}}},
\qquad
a=a_{0}\cdot 10^{i^{\kappa_{a}}},
\qquad
\eta=\eta_{0}\cdot 10^{i^{\kappa_{\eta}}},
\end{equation}
where the constants $E_{0}$, $a_{0}$, $\eta_{0}$ and
$\kappa_{E}$, $\kappa_{a}$, $\kappa_{\eta}$ are found by
the least-squares algorithm.

We proceed by fitting observations for a natural rubber vulcanizate
with an unspecified composition in tensile loading--unloading tests
with an increasing maximal amplitude of stretching, $k_{\max}$.
First, we approximate experimental data for a virgin specimen and
determine adjustable parameters using the steepest-descent procedure.
Afterwards, we fix the constants $b$ and $\nu_{0}$
and match stress--strain curves measured after retraction with the help
of three adjustable parameters: $E$, $a$ and $\eta$.
Figure~4 demonstrates good agreement between experimental data and
results of numerical simulation with adjustable parameters
collected in Table~2.
The quantities $E$, $a$ and $\eta$ are plotted in Figures~5 to 7
versus the number of cycles $i$.
These figures reveal an acceptable quality of fitting experimental
data by Eq. (45).

To evaluate the influence of annealing (24 h at the temperature
$T=100$~$^{\circ}$C) on adjustable parameters in the constitutive
equations, we approximate observations in tensile tests for three
vulcanizates: A (polysulfide crosslinks), B (monosulfide crosslinks)
and C (carbon--carbon crosslinks) with various amounts of crosslinkers
(sulfur and dicumyl peroxide, respectively).
A detailed description of the composition of specimens in presented
in \cite{HP66}.
For any type of vulcanizates, we begin by fitting the stress--strain
curve for a virgin sample with the minimum content of crosslinker,
$\phi$, and find experimental constants by using
the steepest-descent algorithm.
Afterwards, we fix the quantities $b$ and $\nu_{0}$ and match other
observations by using only 3 adjustable parameters: $E$, $a$ and $\eta$.
Experimental data together with results of numerical simulation
are depicted in Figures~8 to 10 for vulcanizate A,
in Figures~11 to 13 for vulcanizate B
and in Figures~14 and 15 for vulcanizate C.
Experimental constants are collected in Tables~3 to 5.

To compare the effects of thermal recovery and recovery by swelling,
we match experimental data in tensile tests for a virgin specimen
and for the same specimen after retraction
(vulcanizates A, B and C with unspecified compositions),
and repeat the same procedure for a recovered sample
(swolen in benzene for 24 h and dried in vacuum).
Adjustable parameters are found by using the same numerical procedure
as for the specimens annealed at an elevated temperature.
Observations and results of numerical simulation are depicted
in Figures~16 to 18, whereas material constants are listed in Table~6.

\section{Discussion}

Figures~1, 4, 8 to 18 demonstrate fair agreement between experimental
data and results of numerical analysis, which implies that the model
may be applied to fit observations in uniaxial tensile tests with
the axial elongation up to $k_{\max}=8.0$.

Figures 3 and 7 show that the parameter $\eta$ increases with the
number of cycles, $i$.
This conclusion is also confirmed by the data listed in Tables~3 to 6
(results for the first and second stretching).
On the contrary, recovery of specimens (by annealing at an elevated
temperature and by swelling) reduces $\eta$.
This means that the quantity $\eta$ may be treated as a parameter
responsible for material damage at the micro-level.
Because $\eta$ reflects the average end-to-end elongation of
a chain driven by an individual trans--cis transition
(which implies that it may be thought of as an average measure of coiling
of a chain), an increase in $\eta$ under cyclic stretching
is tantamount to an increase in the average size of a globule
formed by a chain.
This picture results in the conclusion that recovery of natural
rubber vulcanizates may be treated as coiling of long chains.
Tables~3 to 5 evidence that this process occurs more intensively
for rubbery polymers with high concentrations of crosslinkers,
and vulcanizates with dicumyl peroxide are annealed more effectively
than those with sulfide crosslinks.
These tables also show that the parameter $\eta$ monotonically
decreases with the growth of the content of crosslinker, $\phi$.
Because the growth of the concentration of chemical crosslinks
results in an increase in the volume fraction of junctions
(which is confirmed by the growth of Young's modulus
with $\phi$), and, as a consequence, a decrease in the average
length of a chain, $N$,
we may conclude that short chains are coiled up relatively more
strongly [the qualification ``relatively'' is important here,
because $\eta$ is multiplied by the number of strands
in a chain, $N$, in Eq. (7)] than longer ones.

Tables~3 to 6 show that at the second stretching
the rate of transition from the flexed to extended conformation of bonds,
$a$, exceeds that for a virgin specimen
(the data for vulcanizate B in Table~6 are the only exception
from this rule).
This observation may also be treated as some kind of material damage:
if we take a bond as a hinge, then straightening two rods linked by
the hinge occurs more easily for a preloaded sample than for a virgin one.
Tables~3 to 5 reveal that $a$ substantially grows with the concentration
of crosslinker, $\phi$.
This increase may be associated with appropriate changes in $\eta$:
an increase in the degree of coiling makes trans--cis transitions easier.
In most cases, annealing of specimens (thermal and by swelling)
results in a decrease in $a$ which is accompanied by a decrease
in $\eta$.
However, for vulcanizate A with high concentrations of crosslinker,
an inverse tendency is observed: despite a decrease in $\eta$, the
rate of transition from the flexed to extended conformation increases,
see Tables 3 and 6.

Young's modulus $E$ strongly increases with the content of crosslinker.
This conclusion is in accord with Eq. (40) which implies that $E$
is proportional to the square of the initial number of chains.
Tables~3 to 6 demonstrate that the elastic modulus for a recovered
specimen exceeds that for a prestrained material.
The growth of $E$ during recovery seems quite natural, provided that
the recovery process is associated with an increase in the concentration
of entanglements.

Unlike experimental data collected in Table~2, Table~1 demonstrates
that cyclic loading results in a pronounced decrease in $a$ and $E$
with the number of cycles, $i$.
Figure~2 reveals that this decrease may be fairly well approximated
by the stretched exponential law.
To explain the difference between the data listed in Tables~1 and 2,
we refer to Figures~5 to 7, which evidence
that $E$ and $a$ decrease with $i$ at small amplitudes of stretching
and strongly increase at large amplitudes.
Changes in adjustable parameters observed in the cyclic test
with an increasing amplitude are similar to those presented
in Table~1 during the first 4 cycles and are analogous to
those presented in Tables~3 to 5 when
the maximum amplitude of stretching, $k_{\max}$, exceeds 4.
One can only speculate about a physical mechanism for this phenomenon.
As a possible explanation, we suppose that cyclic loading affects
the internal structure of a rubber vulcanizate in different ways
at small and large amplitudes of stretching.
Periodic loading with amplitudes which do not exceed some
threshold results in material training similar to that
occurring at isothermal annealing (Table~1).
On the contrary, cyclic stretching with amplitudes exceeding
the threshold strain results in damage of the internal
structure which is reflected by the data collected in Tables~2 to 5.

For each series of tests, the parameters $b$ and $\nu_{0}$ are
found from the condition of the best fit for one stress--strain curve
and remain fixed in matching observations for other curves.
Excellent agreement between results of numerical simulation and
experimental data leads to the conclusion that these parameters
are not affected by mechanical factors.

Table~6 reveals that the differences between the parameters $a$ and $E$
found by fitting observations on recovered specimens at the first and second
stretching are rather small compared with those for virgin samples.
This means that recovery by swelling is equivalent to equilibration
of rubbery polymers.
On the contrary, analogous differences in $\eta$ remain practically constant,
which confirms the main hypothesis of this study that stress--softening
of elastomers may be associated with uncoiling of long chains.

\section{Conclusions}

Constitutive equations have been derived for the isothermal
mechanical response of elastomers at finite strains.
Stress--strain relations are developed using the laws of
thermodynamics and are applied to fit experimental data in
uniaxial tensile tests for several natural rubber vulcanizates.
Fair agreement is demonstrated between experimental data
and results of numerical simulation.

We study the effects of cyclic loading, annealing at elevated temperature
and recovery by swelling on material constants.
The following conclusions are drawn:
\begin{enumerate}
\item
Stress-softening of elastomers is reflected by changes in the
parameter $\eta$ which characterizes a mechanically-induced increase
in the average size of globules formed by long chains.
The quantity $\eta$ is increased under periodic stretching
with large amplitudes (uncoiling of chains)
and is decreased at recovery (by annealing or swelling).

\item
In contrast to conventional models for stress-softening of
rubbery polymers, fitting of observations demonstrates that
Young's modulus is altered relatively weakly.
This implies that damage should be associated rather
with changes in the topology of a temporary network
than with rupture of long chains.

\item
Cyclic straining results in a noticeable increase in the rate of
trans--cis transition of bonds, $a$, with the number of cycles, $i$
(which is fairly well described by the stretched exponential law).
On the contrary, recovery of specimens (by thermal annealing and
swelling) induces a decrease in $a$.

\item
Young's modulus, $E$, and the rate of transformation of bonds
from their flexed to extended conformation, $a$, substantially increase
with the content of crosslinker, $\phi$, whereas the parameter $\eta$
is a decreasing function of $\phi$.

\item
Adjustable parameters change in a similar way during thermal
annealing of specimens at an elevated temperature and during recovery
by swelling.

\item
Cyclic loadings with relatively small and large amplitudes
affect the internal structure of elastomers in different ways.
Periodic straining with large amplitudes causes damage
to the internal structure, whereas the influence of that with
small amplitudes is tantamount to recovery of vulcanizate
and may be thought of as a kind of training of rubbery polymers.

\item
The material parameters $b$ and $\nu_{0}$ which reflect
the kinetics of mechanical\-ly-induced transition of loose
entanglements into the active state are rather robust.
They weakly depend on the intensity of loading,
as well as on the composition of vulcanizates and on the content
of crosslinkers.

\item
Recovery by swelling reduces the increments of Young's modulus
and the rate of trans--cis transition driven by a cycle of loading.
This implies that swelling of a rubber vulcanizate induces
equilibration of its mechanical properties.
\end{enumerate}
The above assertions are obtained in the framework of
the phenomenological model, which implies that they should be treated
with caution.
These conclusions may, however, be taken as rather plausible assumptions
(mechanically-induced uncoiling of chains,
training of an elastomer by sub-threshold periodic stretching,
acceleration of trans--cis transformations under loading,
disentanglement at retraction, etc.)
which have to be checked experimentally.
Validation of these hypotheses will be the subject of a subsequent
study.

\subsubsection*{Acknowledgement}

ADD acknowledges financial support by the Israeli Ministry of Science
through grant 1202--1--00.

\newpage
\section*{Figure legends}

\hspace*{6 mm}
Figure~1: The true stress $\sigma$ MPa versus the engineering strain
$\epsilon$ for natural rubber with an unspecified composition.
Circles: experimental data \cite{HP66}.
Solid lines: results of numerical simulation.
Curve~1: $i=1$;
curve~2: $i=2$;
curve~3: $i=3$;
curve~4: $i=8$

Figure~2: Young's modulus $E$ MPa (unfilled circles)
and the dimensionless parameter $a$ (filled circles)
versus the number of cycles $i$.
Symbols: treatment of observations \cite{HP66}.
Solid lines: approximation of the experimental data by Eq. (45).
Curve~1: $E_{0}=12.3705$, $\kappa_{E}=-0.0716$;
curve~2: $a_{0}=9.2561$, $\kappa_{a}=-0.1219$

Figure~3: The dimensionless parameter $\eta$ versus the number
of cycles $i$.
Cirles: treatment of observations \cite{HP66}.
Solid lines: approximation of the experimental data by Eq. (45)
with $\eta_{0}=4.1975$ and $\kappa_{\eta}=-0.2055$

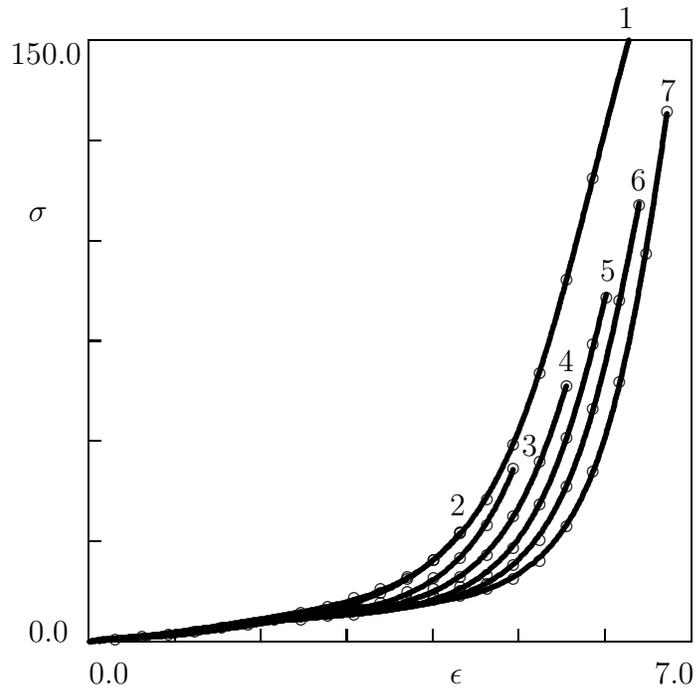
Figure~4: The true stress $\sigma$ MPa versus the engineering strain
$\epsilon$ for natural rubber vulcanizate with an unknown composition.
Circles: experimental data \cite{HP67}.
Solid lines: results of numerical simulation.
Curve~1: a virgin specimen;
curve~2: the 4th stretching;
curve~3: the 5th stretching;
curve~4: the 6th stretching;
curve~5: the 7th stretching;
curve~6: the 8th stretching;
curve~7: the 9th stretching

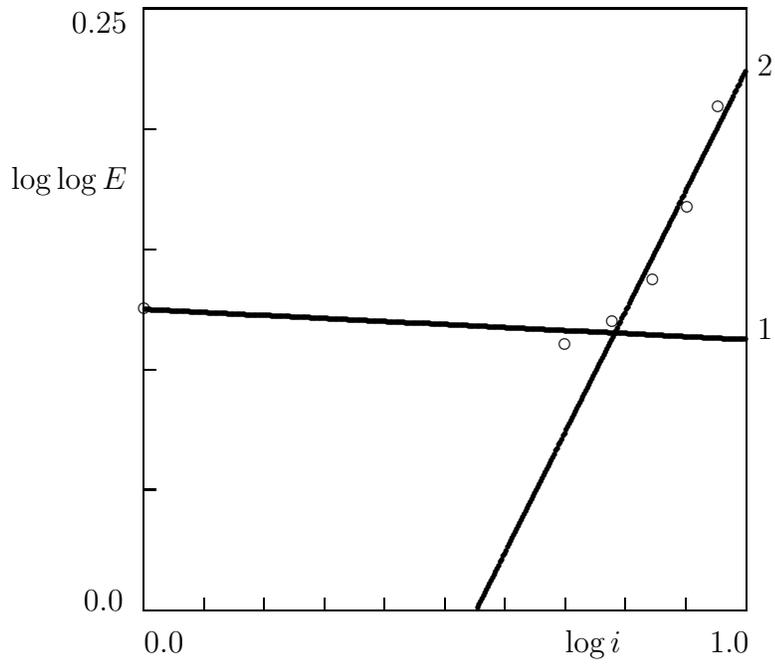
Figure~5: Young's modulus $E$ MPa versus the number of cycles $i$.
Circles: treatment of observations \cite{HP67}.
Solid lines: approximation of the experimental data by Eq. (45).
Curve~1: $E_{0}=21.5638$, $\kappa_{E}=-0.0125$;
curve~2: $E_{0}=3.3807$, $\kappa_{E}=0.5003$

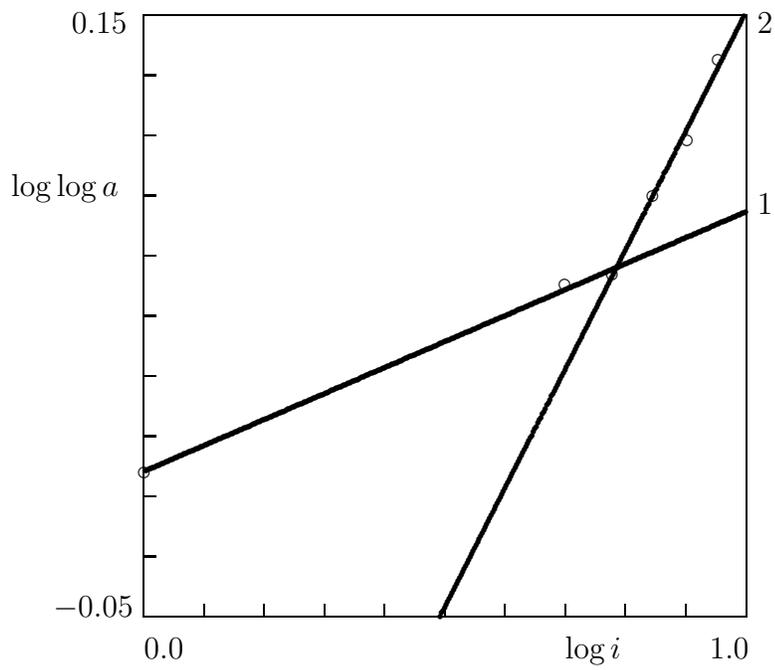
Figure~6: The dimensionless parameter $a$ versus the number of cycles $i$.
Circles: treatment of observations \cite{HP67}.
Solid lines: approximation of the experimental data by Eq. (45).
Curve~1: $a_{0}=9.9095$, $\kappa_{a}=0.0863$;
curve~2: $a_{0}=3.7108$, $\kappa_{a}=0.3952$

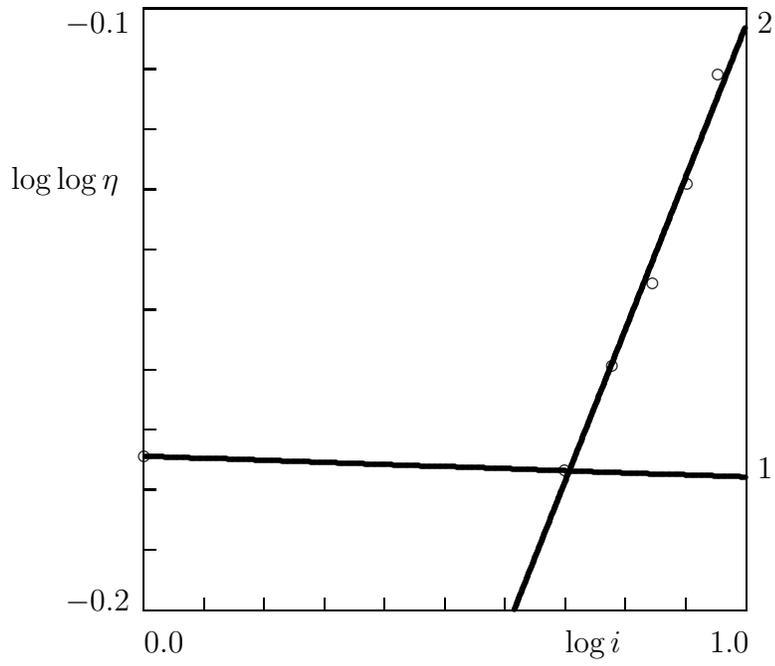
Figure~7: The dimensionless parameter $\eta$ versus the number
of cycles $i$.
Circles: treatment of observations \cite{HP67}.
Solid lines: approximation of the experimental data by Eq. (45).
Curve~1: $\eta_{0}=4.6700$, $\kappa_{\eta}=-0.0035$;
curve~2: $\eta_{0}=2.7620$, $\kappa_{\eta}=0.2524$

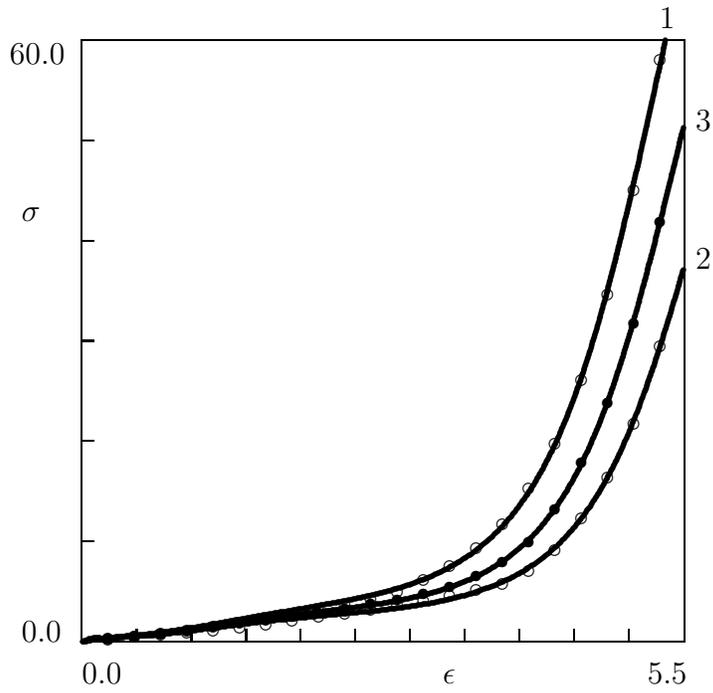
Figure~8: The true stress $\sigma$ MPa versus the engineering strain
$\epsilon$ for natural rubber vulcanizate A ($\phi=1.25$ phr).
Circles: experimental data \cite{HP66}.
Solid lines: results of numerical simulation.
Curve~1: the first stretching;
curve~2: the second stretching;
curve~3: stretching after annealing

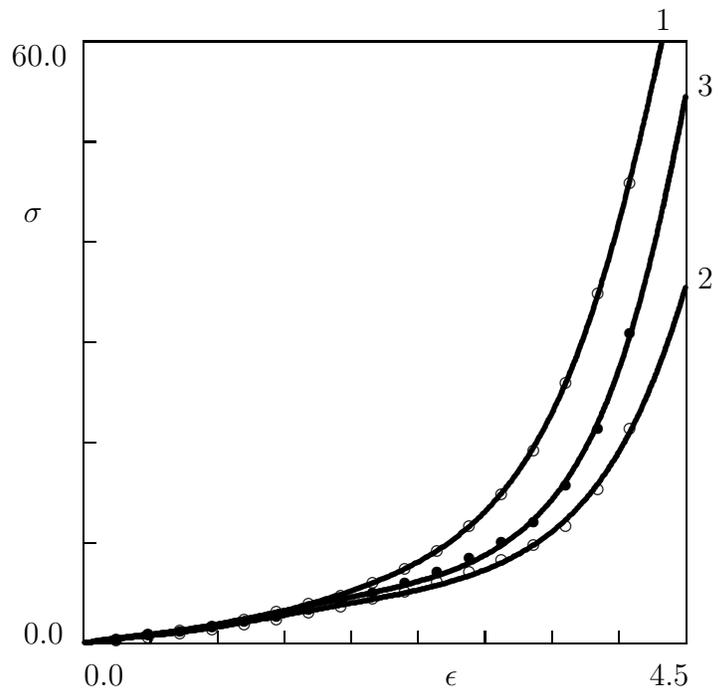
Figure~9: The true stress $\sigma$ MPa versus the engineering strain
$\epsilon$ for natural rubber vulcanizate A ($\phi=2.50$ phr).
Circles: experimental data \cite{HP66}.
Solid lines: results of numerical simulation.
Curve~1: the first stretching;
curve~2: the second stretching;
curve~3: stretching after annealing

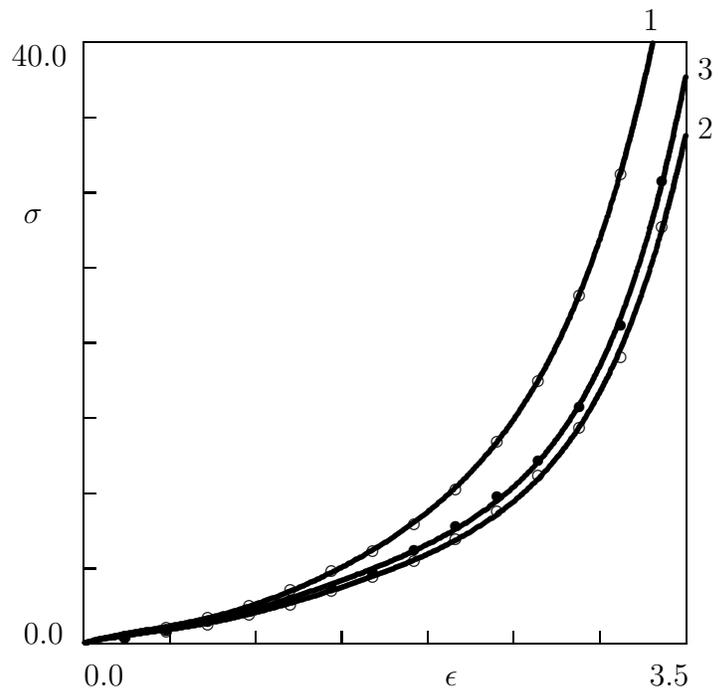
Figure~10: The true stress $\sigma$ MPa versus the engineering strain
$\epsilon$ for natural rubber vulcanizate A ($\phi=4.17$ phr).
Circles: experimental data \cite{HP66}.
Solid lines: results of numerical simulation.
Curve~1: the first stretching;
curve~2: the second stretching;
curve~3: stretching after annealing

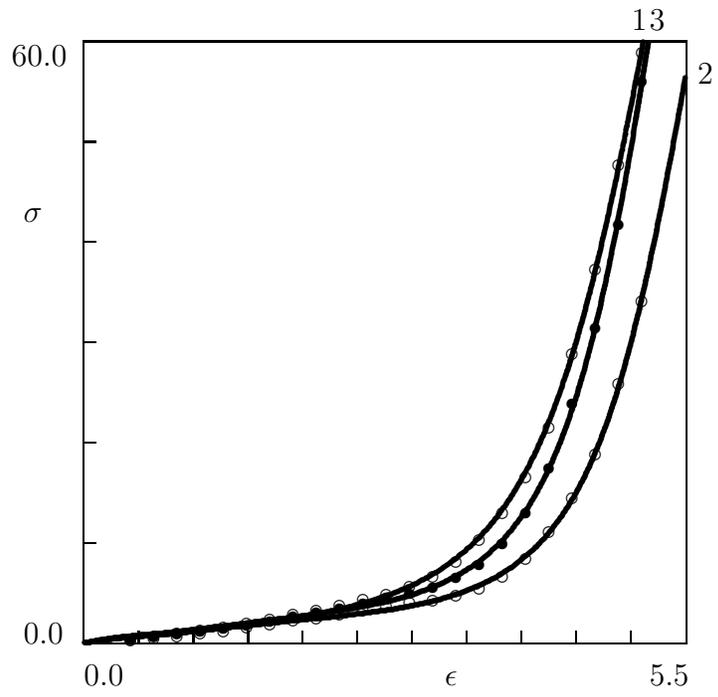
Figure~11: The true stress $\sigma$ MPa versus the engineering strain
$\epsilon$ for natural rubber vulcanizate B ($\phi=0.2$ phr).
Circles: experimental data \cite{HP66}.
Solid lines: results of numerical simulation.
Curve~1: the first stretching;
curve~2: the second stretching;
curve~3: stretching after annealing

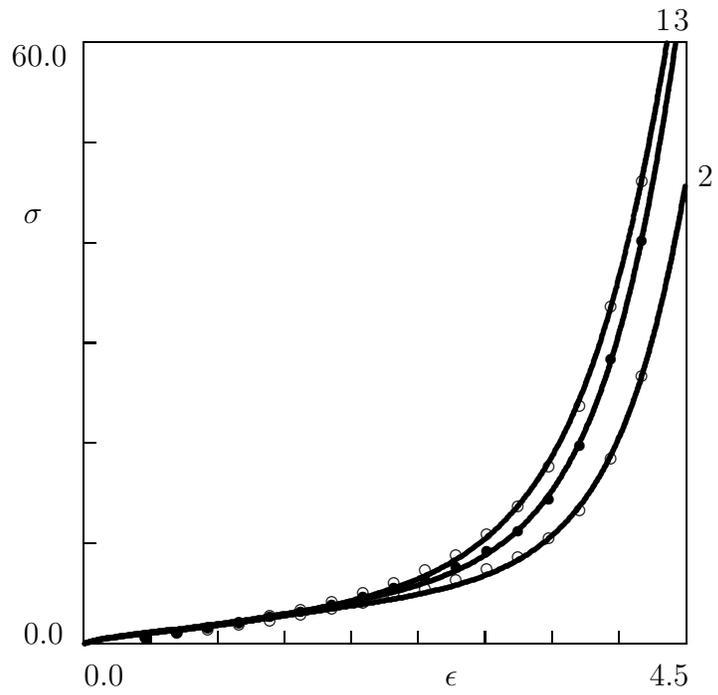
Figure~12: The true stress $\sigma$ MPa versus the engineering strain
$\epsilon$ for natural rubber vulcanizate B ($\phi=0.4$ phr).
Circles: experimental data \cite{HP66}.
Solid lines: results of numerical simulation.
Curve~1: the first stretching;
curve~2: the second stretching;
curve~3: stretching after annealing

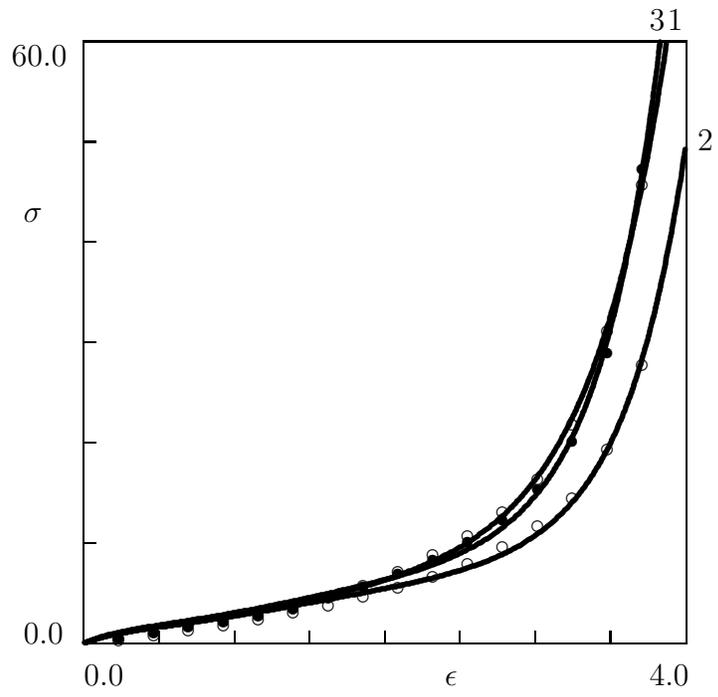
Figure~13: The true stress $\sigma$ MPa versus the engineering strain
$\epsilon$ for natural rubber vulcanizate B ($\phi=0.6$ phr).
Circles: experimental data \cite{HP66}.
Solid lines: results of numerical simulation.
Curve~1: the first stretching;
curve~2: the second stretching;
curve~3: stretching after annealing

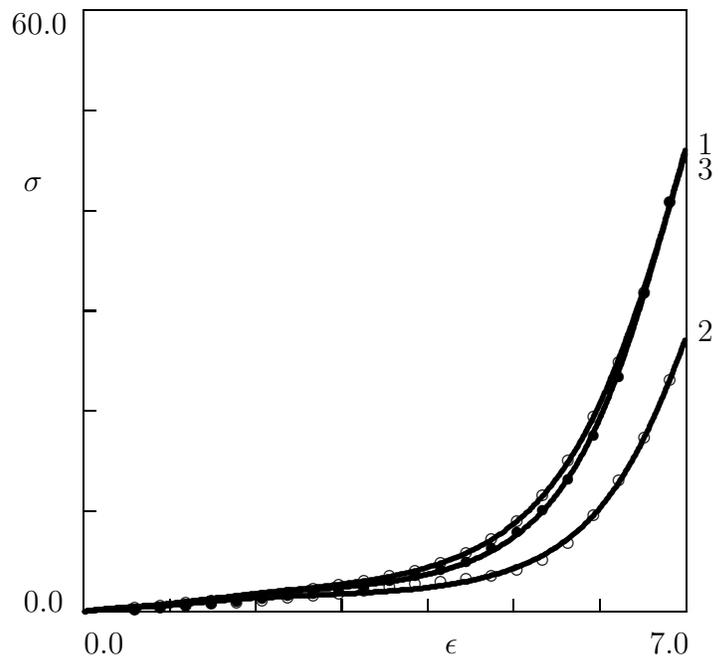
Figure~14: The true stress $\sigma$ MPa versus the engineering strain
$\epsilon$ for natural rubber vulcanizate C ($\phi=0.5$ phr).
Circles: experimental data \cite{HP66}.
Solid lines: results of numerical simulation.
Curve~1: the first stretching;
curve~2: the second stretching;
curve~3: stretching after annealing

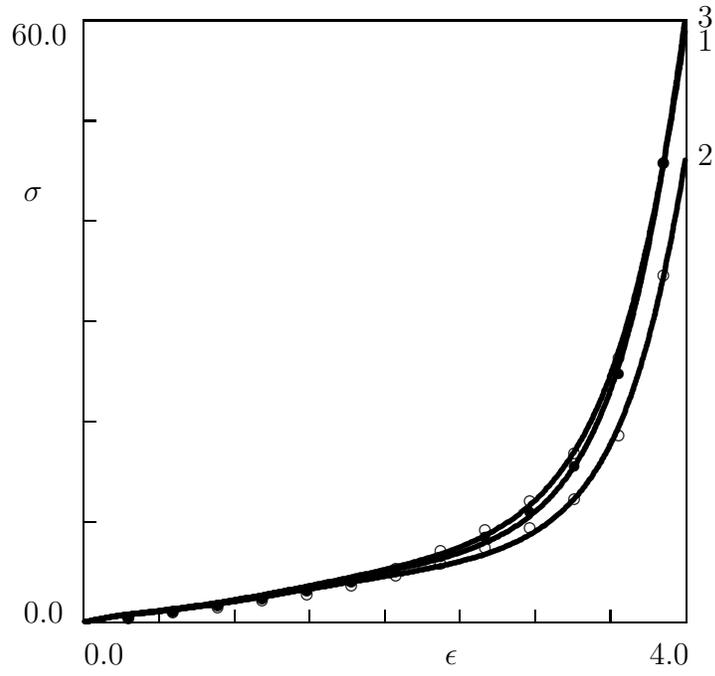
Figure~15: The true stress $\sigma$ MPa versus the engineering strain
$\epsilon$ for natural rubber vulcanizate C ($\phi=3.5$ phr).
Circles: experimental data \cite{HP66}.
Solid lines: results of numerical simulation.
Curve~1: the first stretching;
curve~2: the second stretching;
curve~3: stretching after annealing

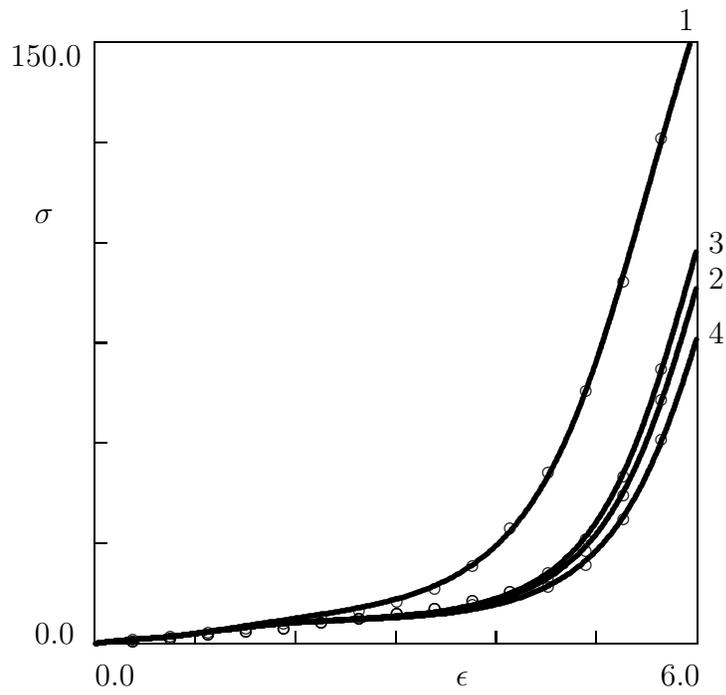
Figure~16: The true stress $\sigma$ MPa versus the engineering strain
$\epsilon$ for natural rubber vulcanizate A with an unspecified
composition.
Circles: experimental data \cite{HP66}.
Solid lines: results of numerical simulation.
Curve~1: a virgin sample, the first stretching;
curve~2: a virgin sample, the second stretching;
curve~3: a recovered sample, the first stretching;
curve~4: a recovered sample, the second stretching

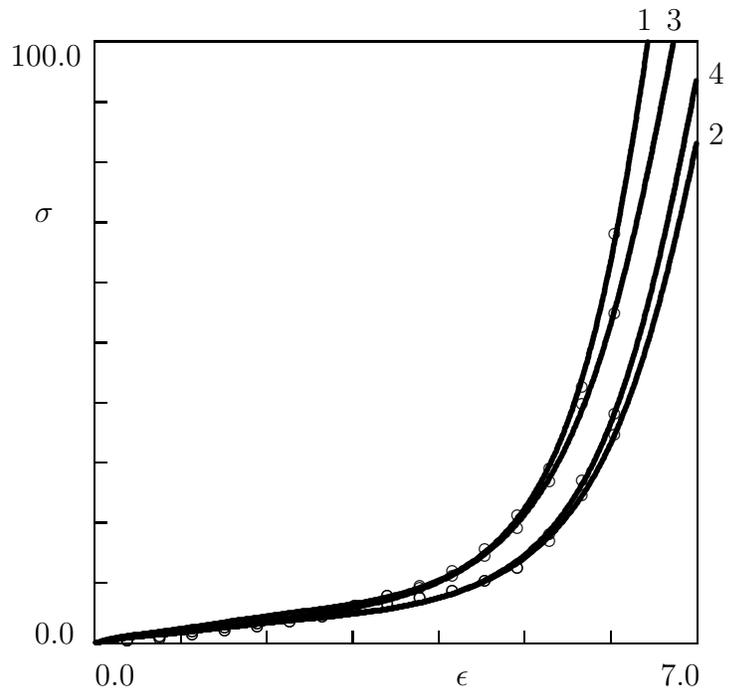
Figure~17: The true stress $\sigma$ MPa versus the engineering strain
$\epsilon$ for natural rubber vulcanizate B with an unspecified
composition.
Circles: experimental data \cite{HP66}.
Solid lines: results of numerical simulation.
Curve~1: a virgin sample, the first stretching;
curve~2: a virgin sample, the second stretching;
curve~3: a recovered sample, the first stretching;
curve~4: a recovered sample, the second stretching

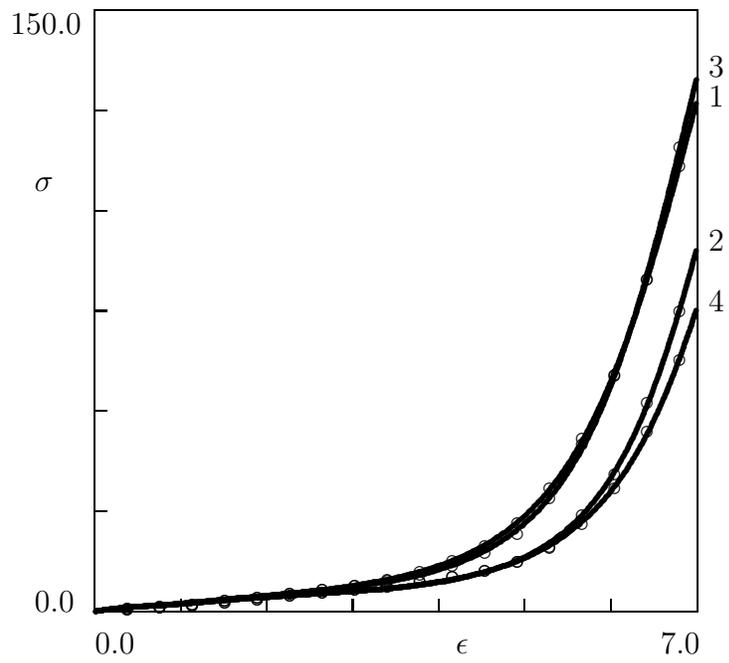
Figure~18: The true stress $\sigma$ MPa versus the engineering strain
$\epsilon$ for natural rubber vulcanizate C with an unspecified
composition.
Circles: experimental data \cite{HP66}.
Solid lines: results of numerical simulation.
Curve~1: a virgin sample, the first stretching;
curve~2: a virgin sample, the second stretching;
curve~3: a recovered sample, the first stretching;
curve~4: a recovered sample, the second stretching

\newpage
\begin{table}[t]
\begin{center}
\caption{Adjustable parameters for NR vulcanizate}
\vspace*{6 mm}

\begin{tabular}{@{} l c c r c c @{}}
\hline
&                $a$ & $\eta$ & $E$ MPa & $\nu_{0}$ & $b$ \cr
\hline
1st stretching   & 8.9  & 4.12 & 12.48 & 1.7 & 0.6\\
2nd stretching   & 8.0  & 4.41 & 10.89 & 1.7 & 0.6\\
3rd stretching   & 7.8  & 4.50 & 10.13 & 1.7 & 0.6\\
8th stretching   & 5.5  & 4.57 &  8.79 & 1.7 & 0.6\\
\hline
\end{tabular}
\end{center}
\end{table}

\begin{table}[b]
\begin{center}
\caption{Adjustable parameters for NR vulcanizate}
\vspace*{6 mm}

\begin{tabular}{@{} l r c r c c @{}}
\hline
&                $a$ & $\eta$ & $E$ MPa & $\nu_{0}$ & $b$ \cr
\hline
1st stretching   & 9.9  & 4.67 & 21.65 & 1.7 & 0.35\\
4th stretching   & 7.9  & 4.36 & 14.69 & 1.7 & 0.35\\
5th stretching   & 14.1 & 4.63 & 19.52 & 1.7 & 0.35\\
6th stretching   & 14.4 & 4.93 & 20.85 & 1.7 & 0.35\\
7th stretching   & 17.0 & 5.19 & 23.61 & 1.7 & 0.35\\
8th stretching   & 19.2 & 5.53 & 29.64 & 1.7 & 0.35\\
9th stretching   & 23.2 & 5.95 & 41.66 & 1.7 & 0.35\\
\hline
\end{tabular}
\end{center}
\end{table}

\begin{table}[t]
\begin{center}
\caption{Adjustable parameters for vulcanizate A}
\vspace*{6 mm}

\begin{tabular}{@{} l c c r c c @{}}
\hline
&                $a$ & $\eta$ & $E$ MPa & $\nu_{0}$ & $b$ \cr
\hline
$\phi =1.25$ phr & & & & &\\
\hline
1st stretching   & 13.5 & 4.27 & 12.66 & 1.6 & 0.4\\
2nd stretching   & 15.1 & 4.67 & 10.25 & 1.6 & 0.4\\
after annealing  & 15.0 & 4.47 & 11.41 & 1.6 & 0.4\\
\hline
$\phi =2.50$ phr & & & & &\\
\hline
1st stretching   & 11.5 & 3.55 & 15.15 & 1.6 & 0.4\\
2nd stretching   & 14.4 & 4.11 & 15.15 & 1.6 & 0.4\\
after annealing  & 18.2 & 3.90 & 17.78 & 1.6 & 0.4\\
\hline
$\phi =4.17$ phr & & & & &\\
\hline
1st stretching   & 12.3 & 3.21 & 22.24 & 1.6 & 0.4\\
2nd stretching   & 17.6 & 3.32 & 18.66 & 1.6 & 0.4\\
after annealing  & 18.4 & 3.32 & 21.06 & 1.6 & 0.4\\
\hline
\end{tabular}
\end{center}
\end{table}

\begin{table}[b]
\begin{center}
\caption{Adjustable parameters for vulcanizate B}
\vspace*{6 mm}

\begin{tabular}{@{} l c c r c c @{}}
\hline
&                $a$ & $\eta$ & $E$ MPa & $\nu_{0}$ & $b$ \cr
\hline
$\phi=0.2$ phr & & & & &\\
\hline
1st stretching   & 10.9 & 4.60 & 13.43 & 1.7 & 0.2\\
2nd stretching   & 16.8 & 5.10 & 14.35 & 1.7 & 0.2\\
after annealing  & 16.4 & 4.81 & 16.04 & 1.7 & 0.2\\
\hline
$\phi=0.4$ phr & & & & &\\
\hline
1st stretching   & 14.4 & 3.90 & 22.69 & 1.7 & 0.2\\
2nd stretching   & 24.7 & 4.20 & 24.26 & 1.7 & 0.2\\
after annealing  & 20.4 & 4.00 & 25.01 & 1.7 & 0.2\\
\hline
$\phi=0.6$ phr & & & & &\\
\hline
1st stretching   & 20.4 & 3.66 & 34.84 & 1.7 & 0.2\\
2nd stretching   & 34.7 & 3.83 & 35.40 & 1.7 & 0.2\\
after annealing  & 33.0 & 3.62 & 39.12 & 1.7 & 0.2\\
\hline
\end{tabular}
\end{center}
\end{table}

\begin{table}[t]
\begin{center}
\caption{Adjustable parameters for vulcanizate C}
\vspace*{6 mm}

\begin{tabular}{@{} l r c r c c @{}}
\hline
&                $a$ & $\eta$ & $E$ MPa & $\nu_{0}$ & $b$ \cr
\hline
$\phi=0.5$ phr &&&&\\
\hline
1st stretching   & 8.4  & 5.61 & 8.37 & 1.7 & 0.3\\
2nd stretching   & 11.7 & 6.02 & 7.01 & 1.7 & 0.3\\
after annealing  & 11.1 & 5.65 & 8.65 & 1.7 & 0.3\\
\hline
$\phi=3.5$ phr &&&&\\
\hline
1st stretching   & 28.2 & 3.58 & 23.33 & 1.7 & 0.3\\
2nd stretching   & 38.1 & 3.67 & 22.34 & 1.7 & 0.3\\
after annealing  & 35.9 & 3.57 & 23.82 & 1.7 & 0.3\\
\hline
\end{tabular}
\end{center}
\end{table}

\begin{table}[b]
\begin{center}
\caption{Adjustable parameters for natural rubber vulcanizates}
\vspace*{6 mm}

%
\end{center}
\vspace*{10 mm}

\caption{}
\end{figure}

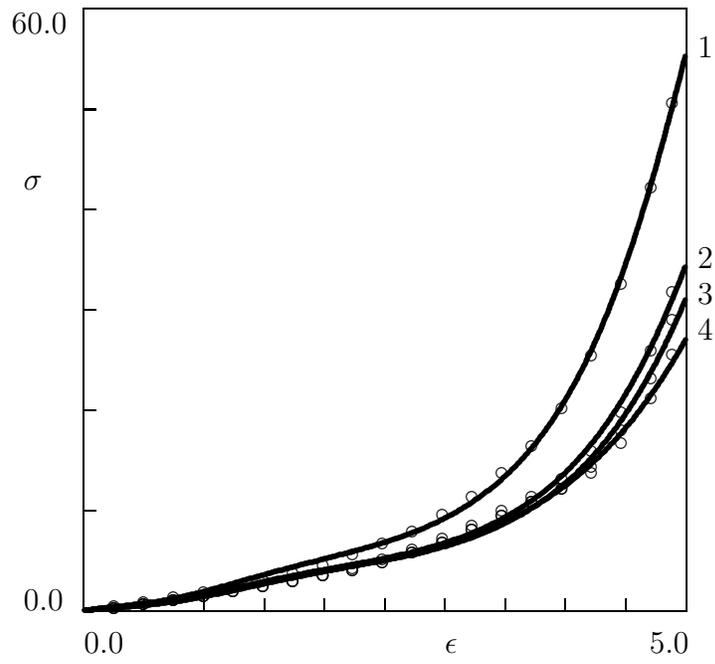
\begin{figure}[t]
\begin{center}
%
\end{center}
\vspace*{10 mm}

\caption{}
\end{figure}

\begin{figure}[t]
\begin{center}
%
\end{center}
\vspace*{10 mm}

\caption{}
\end{figure}

\begin{figure}[t]
\begin{center}
%
\end{center}
\vspace*{10 mm}

\caption{}
\end{figure}
\end{document}